\documentclass[twocolumn,showpacs,showkeys,preprintnumbers,amsmath,amssymb,aps,pre]{revtex4-1}

\usepackage{graphicx}
\usepackage{dcolumn}
\usepackage{bm}
\usepackage{hyperref}

\begin{document}


\title{The Nature of Explosive Percolation Phase Transition}


\author{Liang Tian}
\author{Da-Ning Shi}
\affiliation{College of Science, Nanjing University of Aeronautics
and Astronautics, Nanjing 211106, China}


\date{\today}

\begin{abstract}
In this Letter, we show that the explosive percolation is a novel
continuous phase transition. The order-parameter-distribution
histogram at the percolation threshold is studied in
Erd\H{o}s-R\'{e}nyi networks, scale-free networks, and square
lattice. In finite system, two well-defined Gaussian-like peaks
coexist, and the valley between the two peaks is suppressed with the
system size increasing. This finite-size effect always appears in
typical first-order phase transition. However, both of the two peaks
shift to zero point in a power law manner, which indicates the
explosive percolation is continuous in the thermodynamic limit. The
nature of explosive percolation in all the three structures is
belong to this novel continuous phase transition. Various scaling
exponents concerning the order-parameter-distribution are obtained.
\end{abstract}


\maketitle

\section{Introduction}

Percolation~\cite{Stauffer94}, the simplest model presenting
continuous phase transition, is one of the fundamental problems in
statistical physics, since it provides deeper understanding of many
other issues through Fortuin-Kasteleyn representation~\cite{FK72}.
The percolation model itself has applications to a wide variety of
different systems, ranging from sol-gel transition and
polymerization~\cite{solgel94,Ziff82}, to conductivity of composite
materials and flow through porous media~\cite{Sahimi94,Andrade00},
to spreading process and robustness in
networks~\cite{newman00,PV01,Huberman,barabasi00,cohen00,Moreira09}.
Hitherto the critical properties in most of these systems are well
described by the universality of percolation model in corresponding
dimensionality.

Strikingly, Achlioptas, D'Souza, and Spencer~\cite{optas09} reported
that the percolation transition for the Erd\H{o}s-R\'{e}nyi (ER)
model~\cite{ER60} may become discontinuous, through a modified
growth procedure known as product rule (PR). They found at the
percolation threshold an abrupt jump in the size of the largest
component, which was named as \emph{explosive percolation} (EP)
compared with the traditional continuous percolation transition. In
light of this, subsequent studies were devoted to uncovering the
underlying mechanism of EP~\cite{Friedman09}, proposing new models
for EP~\cite{Moreira10,Souza10,arujo10,cho1pre10,cho2pre10,manna09},
and studying EP transition with different topologies and
dimensionalities~\cite{Ziff09,Ziffpre,choprl09,fortuprl09,fortupre10}.
Recently, two empirical studies focused on the EP in human protein
network~\cite{rozenfeld10} and social network~\cite{raj10}.

While further investigations confirmed the abrupt transition in EP,
it was also shown that the critical distribution of cluster sizes
follows a power law~\cite{fortupre10}, which manifests the features
characteristic of the second-order phase transition. The
contradictions make the nature of explosive percolation transition a
controversial issue in statistical physics, which needs to be
clarified. Recent research proves that the explosive percolation is
a \emph{weak} continuous phase transition in mean-field
structure~\cite{costa10,peter11,oliver11,kim11}. It is reminiscent
of the \emph{weak} first-order phase transition~\cite{nienhuis79} in
five-state Potts model~\cite{wu82}, where, since the correlation
length is very large at transition point, the accessible system size
in numerical simulation is always in the critical region, and thus
the picture of cluster distribution is characterized by fractal
shapes rather than smooth droplets. The weak first-order phase
transition is hard to establish due to its proximity and resemblance
to a critical point, and weak continuous phase transition is hard to
confirmed due to the smallness of the critical exponent for the
order parameter.

However, in this Letter, we show that the explosive percolation is
not a traditional \emph{weak} continuous phase transition by
examining the distribution histogram of the order parameter $G$
defined as the fraction of vertices in the largest cluster. Three
structures are considered, such as ER network, scale-free (SF)
network, and two-dimensional (2D) lattice. Three key features are
observed in all these structures. Firstly, we find that in finite
system two well-defined Gaussian-like peaks coexist in the
order-parameter-distribution histogram at the percolation threshold,
which represent the nonpercolative phase and percolative phase,
respectively. Secondly, the probability of realizing a configuration
in the intermediate phase between the two peaks is suppressed as a
power law with the system size increasing. Finally, both of the two
peaks shift to zero point in power law manner. These observations
indicate the explosive percolation is a continuous phase transition
with first-order-like finite-size effect. Various scaling exponents
concerning order-parameter-distribution are obtained.

\begin{figure}
\scalebox{0.2}[0.2] {\includegraphics{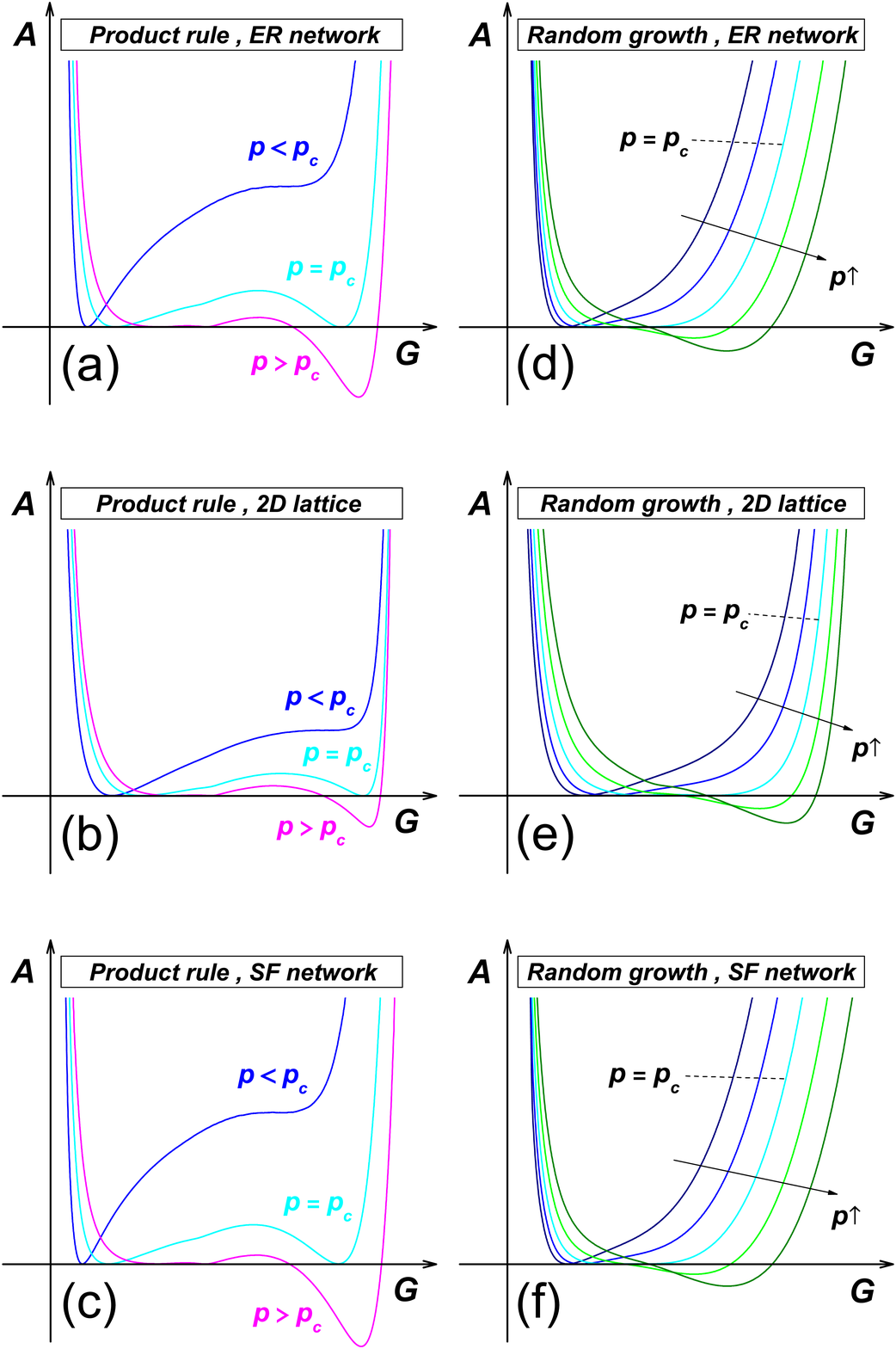}} \caption{Plots of
$A(G,p)$ as a function of order parameter $G$ in the vicinity of
percolation threshold $p_c$ for PR model (a) (b) (c) and traditional
RG model (d) (e) (f). The exponent $\lambda=2.8$ is used for SF
network. The system size is 4096 for all the simulations in this
figure. The full information of $p$ can be found in the movies of
order parameter distribution in the vicinity of transition
point~\cite{tian}.} \label{f.1}
\end{figure}

\section{Method}

For concreteness, numerical simulations were performed according to
the original PR process~\cite{optas09}: In each turn, two unoccupied
edges are randomly chosen; the one which minimizes the product of
the masses of the clusters it joins is retained. For square lattice,
we imposed periodic boundary conditions in both directions to reduce
the boundary effect. For SF network, we adopted the model by Chung
and Lu (CL)~\cite{cl02} to build the network. Specifically, every
vertex in the system is assigned a weight beforehand according to
the desired degree distribution, and at every time step, two edges
are independently selected with probability proportional to the
product of the weights of the vertices at the end of each edge. Then
the PR is used to decide which is the next occupied edge.

The controlling parameter $p$ denotes the number of added edges
divided by the system size $N$. We measured the
order-parameter-distribution histogram $H(G,p)$ for each $p$ through
extensive Monte Carlo (MC) simulations. According to the standard
probability theory, the number of realized configurations with order
parameter $G$ is
\begin{equation}\label{eq1}
    H(G,p) = \exp [-A(G,p)] \sim Z^{-1}(p) Q(G,p),
\end{equation}
where $Z(p)$ is the normalization factor and $Q(G,p)$ is the
order-parameter probability density function, i.e., the probability
that, after $pN$ edges are added with PR process, the fraction of
vertices in the largest cluster is $G$. When the number of
realizations increases to infinity, $H(G,p)$ is identical to
$Q(G,p)$ multiplied by a constant. Intuitively, we have $A(G,p)=-\ln
H(G,p)$, and thus at a given $p$ the location of the global minimum
in $A(G,p)$ denotes the most probable size of the giant component.

\begin{figure}
\scalebox{0.27}[0.27] {\includegraphics{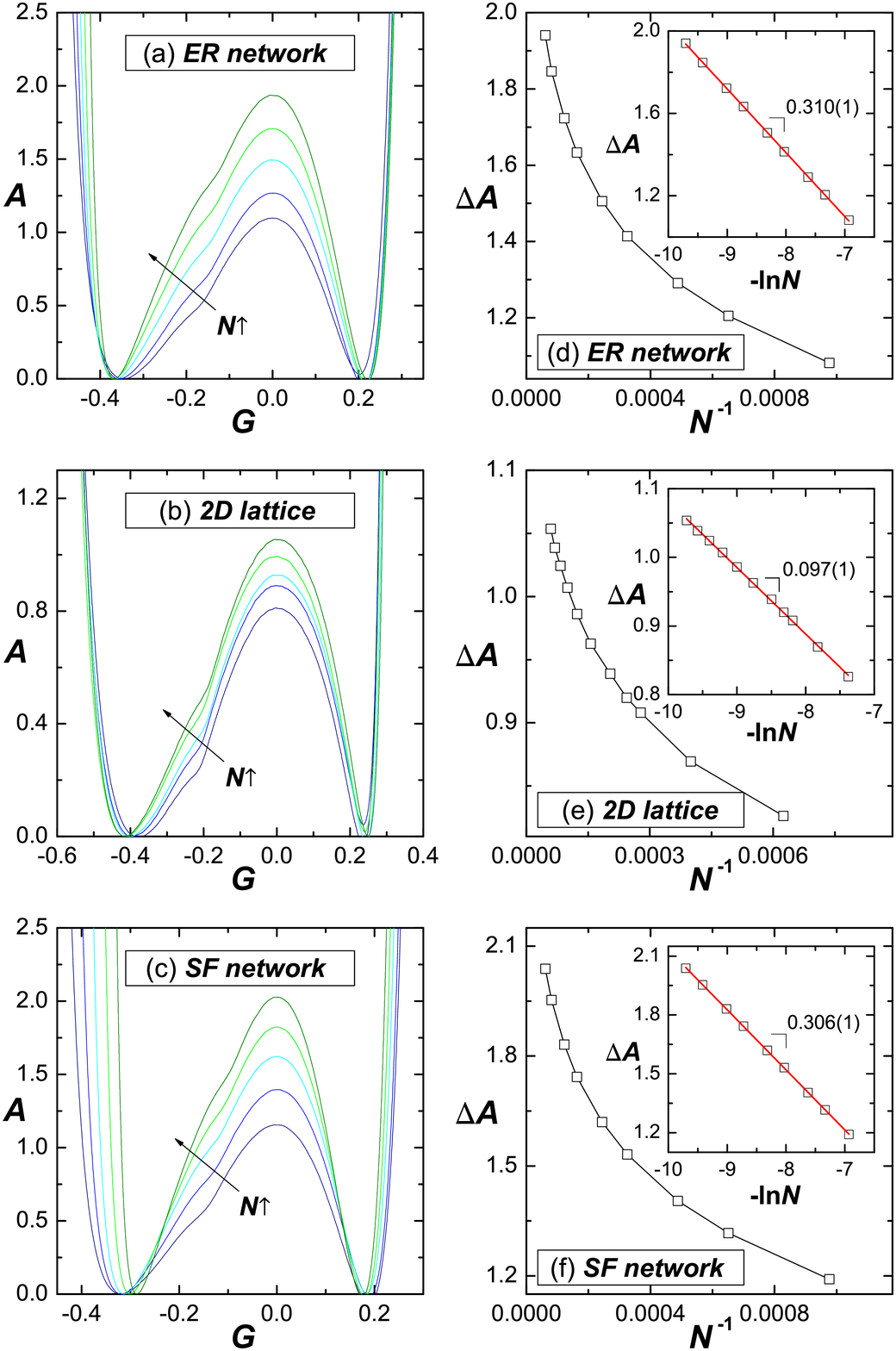}} \caption{The
dependence of $A(G,p)$ at percolation threshold on the order
parameter $G$ for different system sizes (left panels). The values
of the minima have been normalized to 0, and the curves have been
translated along horizontal axis for better comparisons. The system
sizes $N=1024, 2048, 4096, 8192, 16384$ for (a) ER network and (c)
SF network, and $N=1600, 2500, 4900, 8100, 16900$ for (b) 2D
lattice. Right panels display the depth of the minima in $A(G,p_c)$
as a function of system size. The red lines in the insets are the
linear fits. The exponent $\lambda=2.8$ is used for SF network.}
\label{f.2}
\end{figure}

\section{Result and discussion}

Our computer implementation makes use of the effective Newman-Ziff
algorithm~\cite{nz00} for tracking the largest cluster in the
system. We carried out $10^6$ MC sweeps per vertex to achieve high
statistical accuracy for $H(G,p)$. In Fig.~\ref{f.1}, we show the
behavior of $A(G,p)$ near the percolation threshold $p_c$ for both
PR model and traditional random growth (RG) model. It is well known
that the percolation transition with RG is continuous, which is
reproduced in the simulations (see Fig.~\ref{f.1} (d) (e) (f)). As
$p$ passes through the percolation threshold, there is only one
global minimum in $A(G,p)$, which implies the order parameter grows
continuously from one phase to the other. For PR model, the
situation is completely different. As $p$ goes below the critical
value, a local minimum appears in the region of large order
parameter, and its value gradually approaches that of the global
one. Right at the percolation threshold $p_c$, the two minima have
equal depth, indicating that the nonpercolative and percolative
configurations are realized with equal probability. When $p$ is
beyond $p_c$, the second minimum becomes global and percolative
phase dominates. The physical picture of the whole
process~\cite{tian} is reminiscent of the Landau theory of
first-order phase transition. This is the first finite-size property
of explosive percolation.

\begin{figure}
\scalebox{0.20}[0.2] {\includegraphics{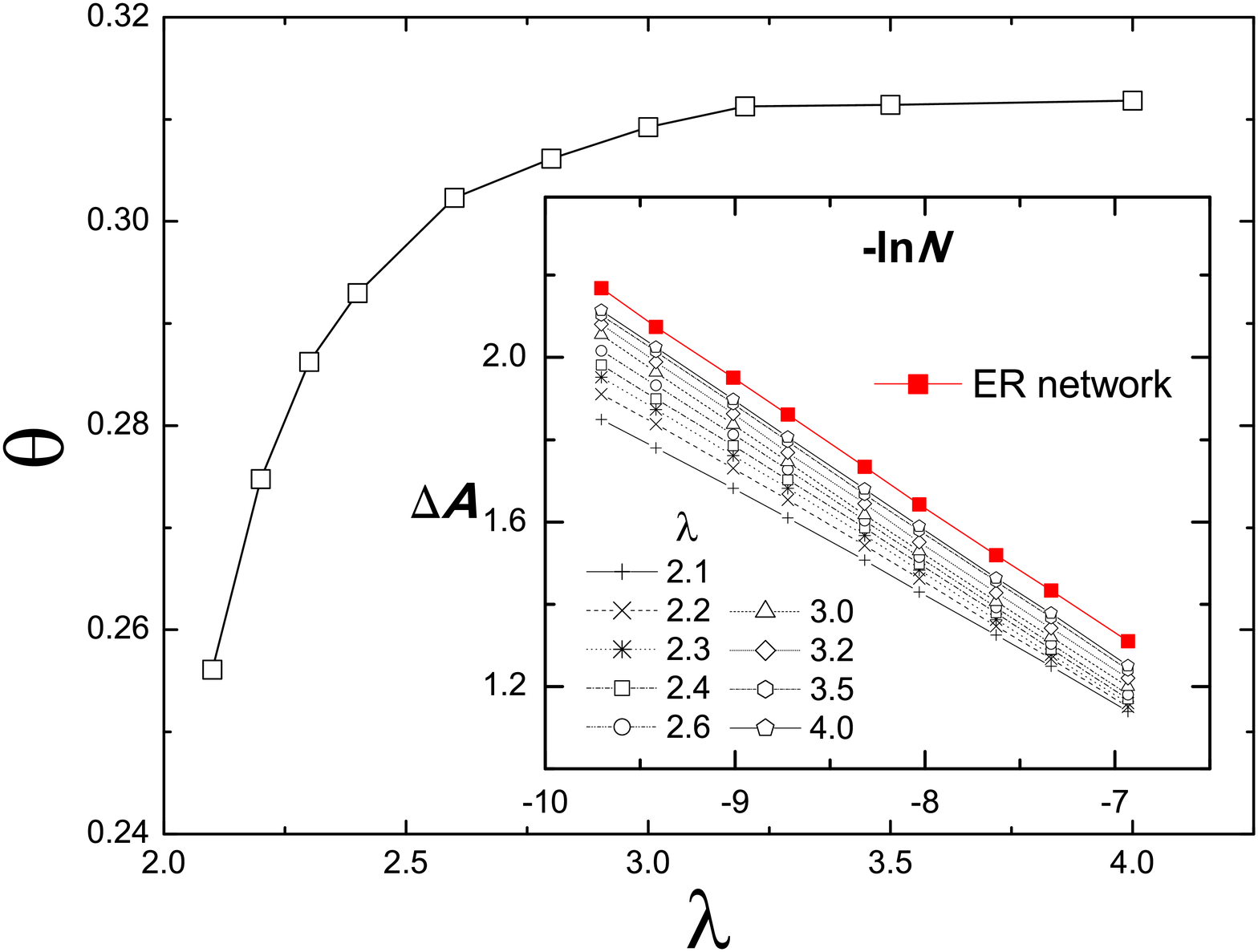}} \caption{The inset
shows in SF network the size-dependent depth of minima in $A(G,p)$
at percolation threshold for different $\lambda$. The red solid
squares represent the same relation in ER network. The fitting
exponent $\theta$ as a function of $\lambda$ is displayed in the
main panel.} \label{f.3}
\end{figure}

In the following, we perform detailed finite-size scaling analysis
of $A(G,p_c)$. An important quantity is the depth of the minima
(corresponding to the peaks in $H(G,P)$), $\Delta A$, relative to
the local maximum (corresponding to the valley in $H(G,P)$)
in-between corresponding to the intermediate phase. For typical
first-order phase transitions~\cite{Binder87,Lee90}, with the system
size increasing, $\Delta A$ also monotonically increases as the
minima gradually develop, and eventually goes to infinity in the
thermodynamic limit. We show here that this property also exists in
explosive percolation. Since $p$ denotes the number of added edges
which is not continuous, it is more convenient to calculate the
depth with $\Delta A=A_{max}-(A^1_{min}+A^2_{min})/2$, where
$A_{max}$ is the value of the local maximum and $A^1_{min}$ and
$A^2_{min}$ are those of the two minima. For the determination of
$A^1_{min}$ and $A^2_{min}$, it should be sufficient to use
quadratic fit in the vicinity of the minima corresponding to the
gaussian-like peaks in $H(G,p_c)$. However, $A(G,p_c)$ shows sizable
asymmetry near the minima, thus we use cubic fit to obtain more
accurate results. To determine the value of $A_{max}$, we found it
adequate to use the same fit. Figure~\ref{f.2} shows the simulation
results for the size-dependent behavior of $A(G,p_c)$. Indeed, the
depth of the minima monotonically increases with system size, and
tends towards infinity in the thermodynamic limit. Furthermore, a
clear relation between $\Delta A$ and the logarithm of $N$ is
observed,
\begin{equation}\label{eq2}
    \Delta A \sim \theta \ln N.
\end{equation}
In other words, the relative probability of finding a configuration
in the intermediate phase is suppressed, as the system size
increases, in power-law manner with exponent $\theta$. It should be
noticed that $\theta_{ER}=0.310(1)$ for ER network and
$\theta_{2D}=0.097(1)$ for square lattice are very different. For SF
network, this scaling relation holds for $\lambda>2.0$ (see the
inset of Fig.~\ref{f.3}). As $\lambda$ increases, the value of
exponent $\theta$ gradually approaches that for ER network, and at
$\lambda\rightarrow\infty$ the CL model is identical to ER model. In
fact, $\theta(\lambda)$ is already saturated for $\lambda>3.0$ (see
Fig.~\ref{f.3}), where with PR the SF network generated by CL model
is hardly distinguishable from ER
network~\cite{fortupre10,choprl09}. The exponent $\theta$ for
different structures is listed in Table~\ref{t.1}. This is the
second finite-size property of explosive percolation.

\begin{figure}
\scalebox{0.2}[0.2] {\includegraphics{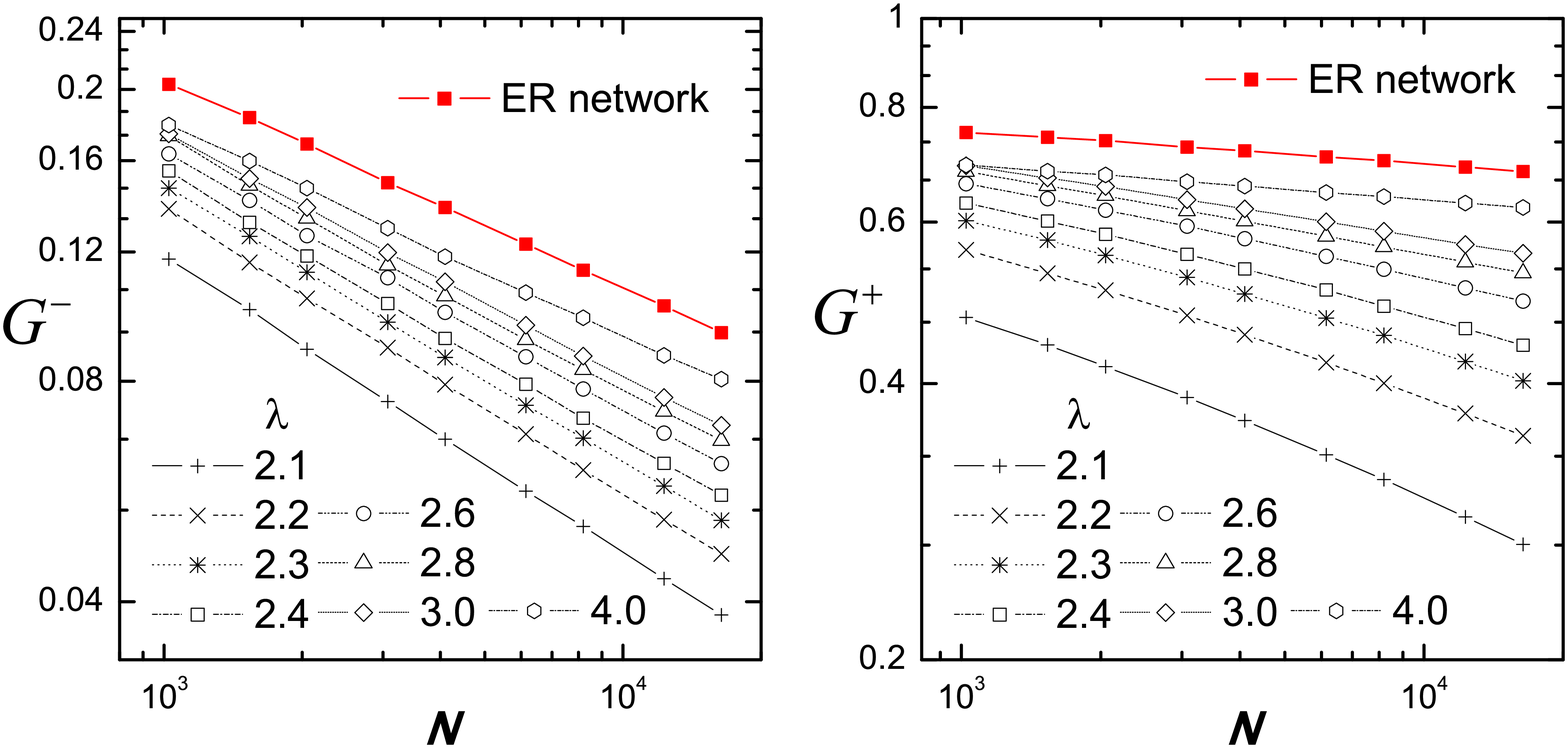}} \caption{The locations
of the two minima $G^-$ (left) and $G^+$ (right) as functions of
system size $N$ for scale-free networks and ER network.} \label{f.4}
\end{figure}

The locations of the two minima in order-parameter dimension, $G^+$
and $G^-$, directly measures the jump of the order parameter at
percolation threshold. In all the three structures, both of $G^+$
and $G^-$ shift to zero in a power law manner, $G^\pm\sim\eta^\pm$.
As $N$ increases, the two minima gradually get close to each other,
and in the thermodynamic limit they merge at the transition point of
the order parameter, where it presents a continuous phase
transition. Actually, in the research of explosive percolation on
scale-free network where another field, the degree distribution
exponent $\lambda$, comes into play, it was claimed that there
exists a tricritical point (TP) at $\lambda_c\in(2.3,2.4)$, above
which the EP transition is first order~\cite{choprl09}. At the same
time, however, careful finite-size scaling analysis implied that for
$\lambda<3.0$ the EP transition is continuous~\cite{fortuprl09}. Our
results indicate that explosive percolation in scale-free network is
continuous in the whole range of $\lambda$. The scaling behavior of
$G^\pm$ in scale-free networks are displayed in Fig.~\ref{f.4}. As
$\lambda$ increases, the value of exponents $\eta^\pm$ gradually
approaches those for ER network. The exponents $\eta^\pm$ for
different structures are listed in Table~\ref{t.1}.

\begin{table}
\caption{\label{t.1}Summary of scaling exponents obtained from our
numerical analysis.}
\begin{ruledtabular}
\begin{tabular}{lccc}
 System & $\eta^-$ & $\eta^+$ & $\theta$ \\\hline
 SF $\lambda=2.1$ & 0.404(1) & 0.205(3) & 0.256(2) \\
 SF $\lambda=2.2$ & 0.389(2) & 0.168(2) & 0.274(1) \\
 SF $\lambda=2.3$ & 0.376(2) & 0.145(1) & 0.286(1) \\
 SF $\lambda=2.4$ & 0.366(2) & 0.129(1) & 0.292(1) \\
 SF $\lambda=2.6$ & 0.351(2) & 0.106(1) & 0.302(1) \\
 SF $\lambda=2.8$ & 0.343(2) & 0.091(1) & 0.306(1) \\
 SF $\lambda=3.0$ & 0.331(2) & 0.079(1) & 0.309(1) \\
 SF $\lambda=3.2$ & 0.293(1) & 0.047(1) & 0.311(1) \\
 SF $\lambda=3.5$ & 0.291(2) & 0.041(1) & 0.311(1) \\
 SF $\lambda=4.0$ & 0.290(2) & 0.038(1) & 0.311(1) \\
 ER network & 0.282(1) & 0.035(1) & 0.310(1) \\
 2D lattice & 0.085(3) & 0.013(1) & 0.097(1) \\
\end{tabular}
\end{ruledtabular}
\end{table}

\section{conclusion}

In this Letter, we show that the explosive percolation is a novel
continuous phase transition in ER network, scale-free network, and
2D lattice. By examining the order-parameter-distribution histogram
at percolation threshold, it is found that two well-defined
Gaussian-like peaks coexist, which represent the nonpercolative
phase and percolative phase, respectively. Moreover, the probability
of realizing a configuration in the intermediate phase between the
two peaks is suppressed as a power law with the system size
increasing. On the other hand, two peaks gradually get close to each
other with the system size increasing, and in the thermodynamic
limit they merge at the transition point of the order parameter.
These observations suggest the explosive percolation is a continuous
phase transition with first-order-like finite-size effect.

\begin{acknowledgments}
The author (L. T.) thanks Robert Ziff, Peter Grassberger, Bernardo
Huberman, and Lei-Han Tang for stimulating discussions and
suggestions. The financial supports from CX07B-033z and BCXJ07-11
are acknowledged.
\end{acknowledgments}

\bibliography{explosive}

\end{document}